\documentstyle[12pt,psfig,aasms4]{article}

\def\secondip{\hbox{\rlap{\hbox{.}}\hbox{$''$}}}

\def\gradip{\hbox{\rlap{\hbox{.}}\raise 5.truept \hbox{{\small $\circ$}}}}

\catcode`\@=11
\def\gsim{\ifmmode{\mathrel{\mathpalette\@versim>}}
    \else{$\mathrel{\mathpalette\@versim>}$}\fi}
\def\lsim{\ifmmode{\mathrel{\mathpalette\@versim<}}
    \else{$\mathrel{\mathpalette\@versim<}$}\fi}
\def\@versim#1#2{\lower 2.9truept \vbox{\baselineskip 0pt \lineskip 
    0.5truept \ialign{$\m@th#1\hfil##\hfil$\crcr#2\crcr\sim\crcr}}}
\catcode`\@=12
\def\msun{\hbox{$M_\odot$}}
\def\lsun{\hbox{$L_\odot$}}

\begin{document}
\title{The Initial Mass Function of the Galactic Bulge \\ Down to $\sim 0.15\,\msun$
\altaffilmark{1}}
\author{
Manuela Zoccali\altaffilmark{2,3},
Santi Cassisi\altaffilmark{4},
Jay A. Frogel\altaffilmark{5},
Andrew Gould\altaffilmark{5},
Sergio Ortolani\altaffilmark{2,3},
Alvio Renzini\altaffilmark{3},
R. Michael Rich\altaffilmark{6},
Andrew W. Stephens\altaffilmark{5}
}

\altaffiltext{1}{
Based on observations with the NASA/ESA Hubble Space Telescope obtained
at the Space Telescope Science Institute, which is operated by the
Association of Universities for Research in Astronomy, Incorporated, under
NASA contract NAS5-26555.}
\altaffiltext{2}{Dipartimento di Astronomia, Universit\`a di Padova, vicolo 
dell'Osservatorio 5, I-35122 -- Padova -- Italy; zoccali@pd.astro.it, 
ortolani@pd.astro.it}
\altaffiltext{3}{European Southern Observatory, Kark-Schwarzschild-Strasse 2,
D-85748, Garching bei M\"{u}nchen, Germany; arenzini@eso.org}
\altaffiltext{4}{Osservatorio Astronomico di Collurania, Via M. Maggini,
64100 Teramo; cassisi@astrte.te.astro.it}
\altaffiltext{5}{Department of Astronomy, Ohio State University, 5040 Smith 
Lab., 174 West 18th Avenue, Columbus, OH 43210; frogel@astronomy.ohio-state.edu,
gould@astronomy.ohio-state.edu, stephens@astronomy.ohio-state.edu}
\altaffiltext{6}{Department of Physics and Astronomy, Division of Astronomy 
and Astrophysics, University of California, Los Angeles, CA 90095-1562; 
rmr@astro.ucla.edu}


\begin{abstract}

We present a luminosity function (LF) for lower main sequence stars in
the Galactic bulge near $(l,b)=(0^\circ,-6^\circ)$ to $J=24 $,
corresponding to $M_J\sim9.3$.  This LF is derived from Hubble Space
Telescope (HST) + Near Infrared Camera and Multi Object Spectrometer
(NICMOS) observations of a region of $22\secondip 5\times 22\secondip
5$, with the F110W and F160W filters. The main sequence locus in the
infrared shows a strong change in slope at $J\sim20.5$ ($M_{\rm
J}\sim5.75)$ which is well fit by new low mass models that include
water and molecular hydrogen opacity.  Our derived
mass function (which is not corrected for binary companions) is
the deepest measured to date in the bulge, and extends to 0.15
$M_\odot$ with a power law slope of $\alpha=-1.33\pm0.07$; a Salpeter
mass function would have $\alpha=-2.35$.  We also combine our $J$ band
LF with previously published data for the evolved stars to produce a
bulge LF spanning $\sim$ 15 magnitudes.  We show that this mass
function has negligible dependence on the adopted bulge metallicity
and distance modulus.  Although shallower than the Salpeter slope, the
slope of the bulge IMF is steeper than that recently found for the
Galactic disk ($\alpha=-0.8$ and $\alpha=-0.54$ from the data of Reid
\& Gizis, 1997, and Gould et al. 1997, respectively, in the same mass
interval), but is virtually identical to the disk IMF derived by
Kroupa et al. (1993).  The bulge IMF is also quite similar to the mass
functions derived for those globular clusters which are believed to
have experienced little or no dynamical evolution.  Finally, we derive
the $M/L_J$ ratio of the bulge to be $\sim 0.9\pm0.1$ $M_\odot/L_\odot$, 
and briefly discuss the implications of this bulge IMF for the
interpretation of the microlensing events observed in the direction of
the Galactic bulge.

\end{abstract}


\section{Introduction}
\label{intro}

The initial mass function (IMF) is a fundamental property of stellar
populations, hence one of the most crucial ingredients in models of
galaxy formation and evolution. It determines several key properties
of stellar populations and galaxies, such as the yield of heavy
element production, the luminosity evolution over time, the
mass-to-light ratio, the total star formation rate at low and high
redshifts as inferred from empirical estimators, and the energetic
feedback into the interstellar medium. Yet, the IMF is usually taken
as a free parameter, particularly at the low-mass end (for recent
reviews on the IMF see Larson 1998; Scalo 1998, 1999).  Observational
constraints on the IMF are therefore of the greatest astrophysical
importance.

Knowing the IMF at $M \lsim 1 M_\odot$ in spiral bulges and elliptical
galaxies is of special interest because these spheroids contain a
large fraction, perhaps a majority, of all the stellar mass of the
universe (e.g., Fugugita, Hogan \& Peebles, 1998).  However, there is
presently no way to directly determine the IMF of spheroids except by
measuring the luminosity function (LF) of our own bulge as the only
surrogate for the unresolvable population in other galaxies. Although
the low mass end of the stellar IMF has been determined for the solar
neighborhood (Kroupa, Tout \& Gilmore 1993; Gould, Bahcall \& Flynn
1997; Reid \& Gizis 1997) and in young open clusters (Hillenbrand
1997; Bouvier et al. 1998; Luhman et al. 1998) it is only in the
Galactic bulge that one can be confident that the stellar population
is old, largely coeval, and metal rich (Whitford 1978; Ortolani et al. 
1995; Mc William \& Rich 1994), i.e., the closest we can come in a nearby,
resolved stellar population to what prevails in other spiral bulges
and elliptical galaxies (Renzini 1999).

The recent discovery of a high rate of microlensing events towards the
bulge (Udalski et al. 1994; Alcock et al. 1997) has made the
determination of the faint end of the IMF a yet more urgent
problem. In brief, if the bulge IMF is close to that of the
solar-neighborhood (Gould et al. 1997) then the bulk of the short
($\sim 10$ day) microlensing events would remain unexplained, perhaps
requiring a large population of brown dwarfs (Han 1997). However, an
IMF extending to the H-burning limit with a Salpeter law can account
for both the total mass of the bulge and the frequency of microlensing
events (Zhao, Spergel, \& Rich 1995). It is therefore tempting to
suspect that the bulge and solar-neighborhood IMFs are
different.  However, the interpretation of the microlensing events
relies on assumptions about the phase-space distribution of both the
lenses and sources, and some events may be caused by collapsed stars,
brown dwarfs or even non-stellar objects. Therefore, the most reliable
way to resolve these ambiguities (and thus maximize the information
from microlensing itself) is to obtain a representative stellar inventory of
the bulge from star counts, and to incorporate this into the
microlensing analysis.

A recent
determination of the bulge IMF down to $M \sim 0.35 M_\odot$ has been
provided by Holtzman et al. (1998), based on Hubble Space Telescope (HST) + 
Wide Field and Planetary Camera (WFPC2) observations of Baade's Window.
Globular clusters offer another approach towards determining the IMF
for $M \lsim 1M_\odot$, and deep HST observations are indeed providing
important information on their present-day mass functions (De Marchi
et al. 1999; Piotto \& Zoccali 1999, and references therein). However,
clusters suffer from dynamical evolution and evaporation of low-mass
stars, and therefore there is no model-independent way to infer their
IMFs from their observed present day mass function (MF).

Because faint, low-mass stars have such low temperatures, infrared
observations give a crucial advantage over optical data. Moreover, in
the near-IR the effects of extinction and differential reddening are
considerably reduced, and the bolometric luminosities of M dwarfs (the
vast majority of the sampled stars) are best determined in the near-IR
both because of their cool temperatures and severe molecular
blanketing in the optical. Finally, the relatively low IR background
of HST, combined with diffraction limited resolution,
gives a fundamental advantage when dealing with very faint sources in a
crowded field. Therefore, the NICMOS near-IR cameras offer a unique
opportunity to reach the faintest stars possible in the Galactic bulge,
thus extending to lower masses the range over which the IMF is 
observationally constrained.

In order to ensure the success of the project we paid special
attention to the selection of the bulge field to be observed.  The
most widely studied field in the Galactic bulge is the $b=-4^\circ$
field known as Baade's Window.  However, for the NICMOS observations
we did not choose to point HST at this field. A priori, in Baade's
Window, crowding might have been too severe to confidently undertake
this experiment that aims at counting the faintest bulge stars in the
frame. At $b=-4^\circ$ the average surface brightness (corrected for
$A_{\rm V}\simeq 1.2$ mag extinction) is 18.7 $V$ mag$\,$arcsec$^{-2}$
(Terndrup 1988), and with a true modulus of 14.5 mag one samples
$M_{\rm V}=4.2$ mag$\,$arcsec$^{-2}$, corresponding to a bolometric
luminosity of $\sim 2.8 L_\odot\,$arcsec$^{-2}$ (using population
synthesis models, e.g., by Maraston 1998). Hence, the NIC2 camera
samples a total bolometric luminosity $L_{\rm T}\simeq
10^3L_\odot$. This allows one to estimate the number of main sequence
stars in a HST NIC2 frame, knowing that for a $\sim 15$ Gyr old
population the scale factor in the IMF -- $\psi(M)= AM^{\alpha}$ -- is
given by $A\simeq 1.2 L_{\rm T}$ (Renzini 1998).  Integrating the IMF
from 0.1 to 0.9 $M_\odot$, with $\alpha=-2.35$ (the Salpeter IMF
slope), and $A=1.2\times 10^3$, we get that a NIC2 frame will contain
$\sim 2.3\times 10^4$ stars. Since the NIC2 camera has $6.55\times 10^4$
pixels, while selecting the target field we therefore concluded that
accurate photometry would have been difficult towards the faint
end of the LF, if the IMF were to follow the Salpeter's slope all the
way to the hydrogen burning limit.

For our observations we selected instead the field at $b=-6^\circ$,
where the surface brightness is $\sim 1$ mag lower, and then we expect
$\sim 2.5$ times fewer stars in a NIC2 frame compared to Baade's
Window, significantly improving the stars/pixel number ratio.
Although more distant from the nucleus than Baade's Window, the field
population is still dominated by the metal rich stars characteristic
of the bulge, as shown by the strongly descending red giant branch in
the $(V,V-I)$ diagram (Rich et al. 1998), which makes sure that we are
properly studying the metal rich bulge in this location.  Moreover,
photometry down to the hydrogen burning limit should not be
compromised by crowding, especially if the IMF were to flatten out
below the Salpeter's slope as in the solar neighborhood (Gould et
al. 1997), implying a smaller number of low-mass stars.


\section{Observations and Data Analysis}
\label{phot}

\begin{figure}
\vskip 0.4cm
\centerline{\psfig{file=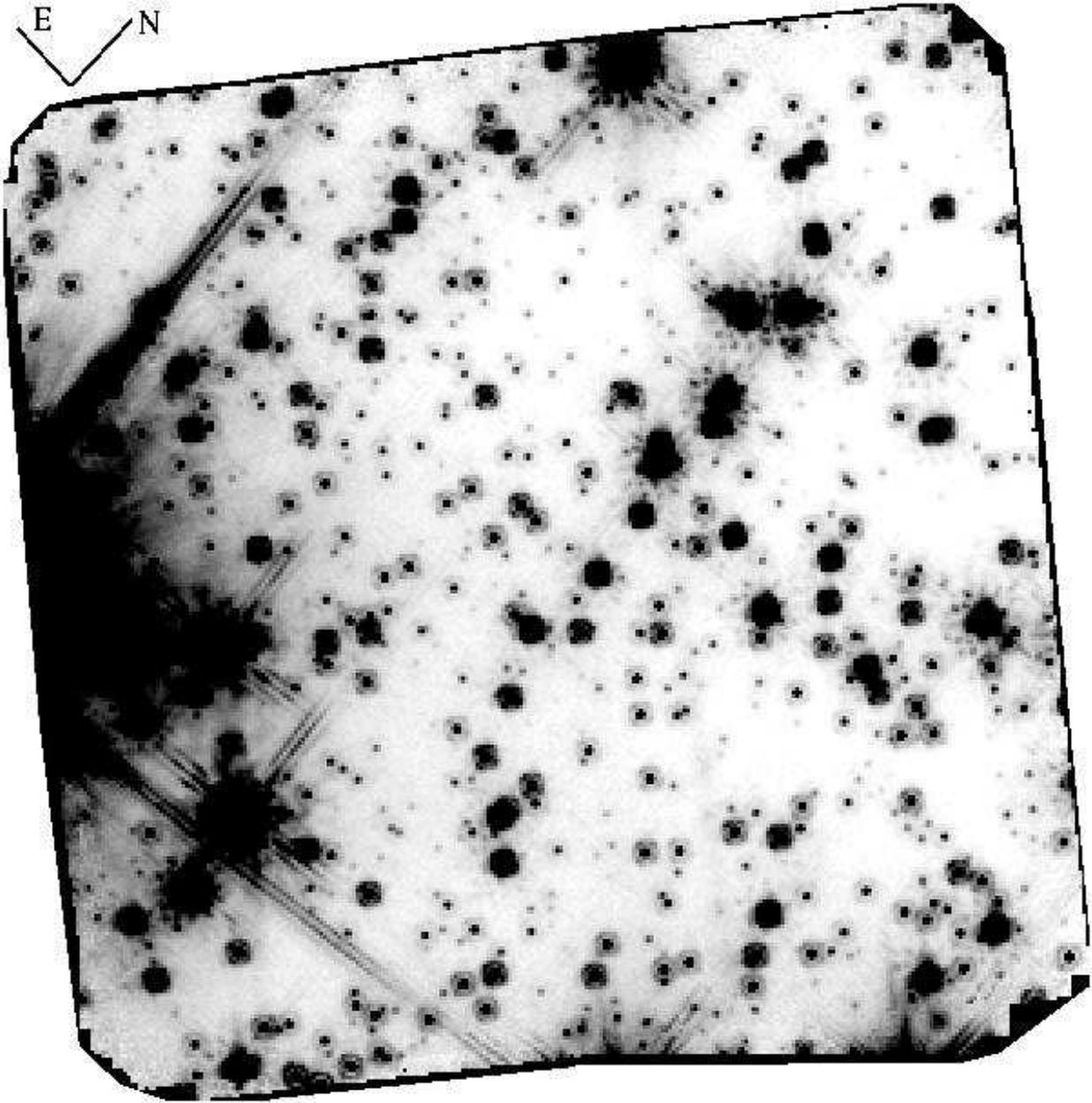,width=17truecm}}
\caption[]{Negative image of the region of the bulge window at $-6^\circ$
observed with the NIC2 camera of NICMOS on board HST . This image is
obtained from the combination of all the F110W and F160W frames. The
total field of view is about $22\secondip 5 \times 22\secondip 5$.}
\label{field}
\end{figure}

The selected field (RA=18:11:05, DEC=$-$31:45:49 (J2000); $l$=0.277, $b$=$-$6.167)
was observed with the NIC2 camera of
NICMOS, on board HST, through the filters F110W and F160W.  Parallel
observations with NIC1 were collected through the F110W filter.
Fourteen orbits were allocated, for total NIC2 integration times of 10240 s
and 25600 s in F110W and F160W, respectively. 
Only the F110W filter was used for the NIC1 parallel observations, for a
total integration time of 35850 s. All exposures were
obtained using the MULTIACCUM readout mode and the STEP64 time sequence
through an eight-position spiral dithering with size of $0\secondip 4$.
The pixel size of the NIC2 detector is $0\secondip 075$, giving a field of
view of $19\secondip 2 \times 19\secondip 2$ for each frame. Small
offsets and rotations among the frames gave us a slightly larger total
field ($22\secondip 5 \times 22\secondip 5$).  Figure~\ref{field} shows
the observed bulge region as it appears in a combination of all the
frames.

The images were bias subtracted, dark corrected and flat-fielded by the
standard NICMOS pipeline CALNICA. This routine also combines the
multiple readouts of the MULTIACCUM mode, giving an output image which
is expressed in counts per second per pixel. We therefore multiplied
each of these images by its total exposure time, so that the photometry
software  would measure the correct signal-to-noise ratio (S/N).

The data quality file corresponding to each image was used to mask out
the saturated and bad pixels by setting them to a very high value,
that was discarded in the photometry. Following the NIC2 manual, the 
read-out noise of each frame was assumed to be 32 electrons, corresponding
to 6.1 ADU, with a conversion factor of 5.4 $e^-/$ADU. The mean sky 
level of each 640 s exposure was $\sim 50$ and $\sim 40$ ADU in F110W
and F160W respectively. That is, the noise is dominated by read noise
rather than the sky. 

Preliminary star finding and aperture photometry was carried out on
each frame using the DAOPHOTII photometry package (Stetson 1987). We
then used all the stars identified in each frame to obtain the
coordinate transformations among all the frames.  These
transformations were used to register the frames and obtain a median
image. 
The latter, having the highest S/N, was used to create the most complete
star list, by means of two complete runs of DAOPHOTII and ALLSTAR. The
final star list, together with the coordinate transformations, was
finally used as input for ALLFRAME (Stetson 1994), for the
simultaneous reduction of all the frames.  Particular attention was
devoted to modeling the NIC2 point spread function (PSF) in the two
filters. This was performed using specific software (MULTIPSF),
provided by P.B.\ Stetson, that allows measurement of a unique PSF
from the brightest and most isolated stars in a set of different
frames.  Assuming that the PSF profile does not change from frame to
frame, we were able to measure the same $\sim30$ stars in all the
frames of each filter.  The spatial dithering allowed us to measure
the selected stars in different locations on the chip, centered in
different positions inside a pixel, so the final PSF was of
considerably better quality than the one we could obtain for each
frame taken individually. The stellar full width at half maximum 
is $\sim 1.5$ pixels while the adopted model PSF was defined up to a 14 
pixels radius.

Aperture corrections were empirically determined on the most isolated
stars, and applied to the ALLFRAME measures in order to obtain the
stellar magnitudes in a $0\secondip 5$ aperture. The magnitudes were
then converted to count rates, and multiplied by 1.15 to correct to an
infinite aperture. The inverse sensitivity, given as the keyword
PHOTFLAM in the header of the images, together with the zero points
PHOTZPT given in Table 2 of Stephens et al. (1999), were used to
convert the count rates into HST $m_{110}$ and $m_{160}$
magnitudes. The latter were then transformed in the CIT/CTIO system
according to the calibration equations determined by Stephens et
al. (1999) from the comparison between NICMOS and ground-based
observations of the same 14 bright stars.  As discussed by Stephens et
al. (1999), this calibration is consistent (over the common color
range) with the one described in the NICMOS calibration documentation.

\begin{figure}
\vskip 0.4cm
\centerline{\psfig{file=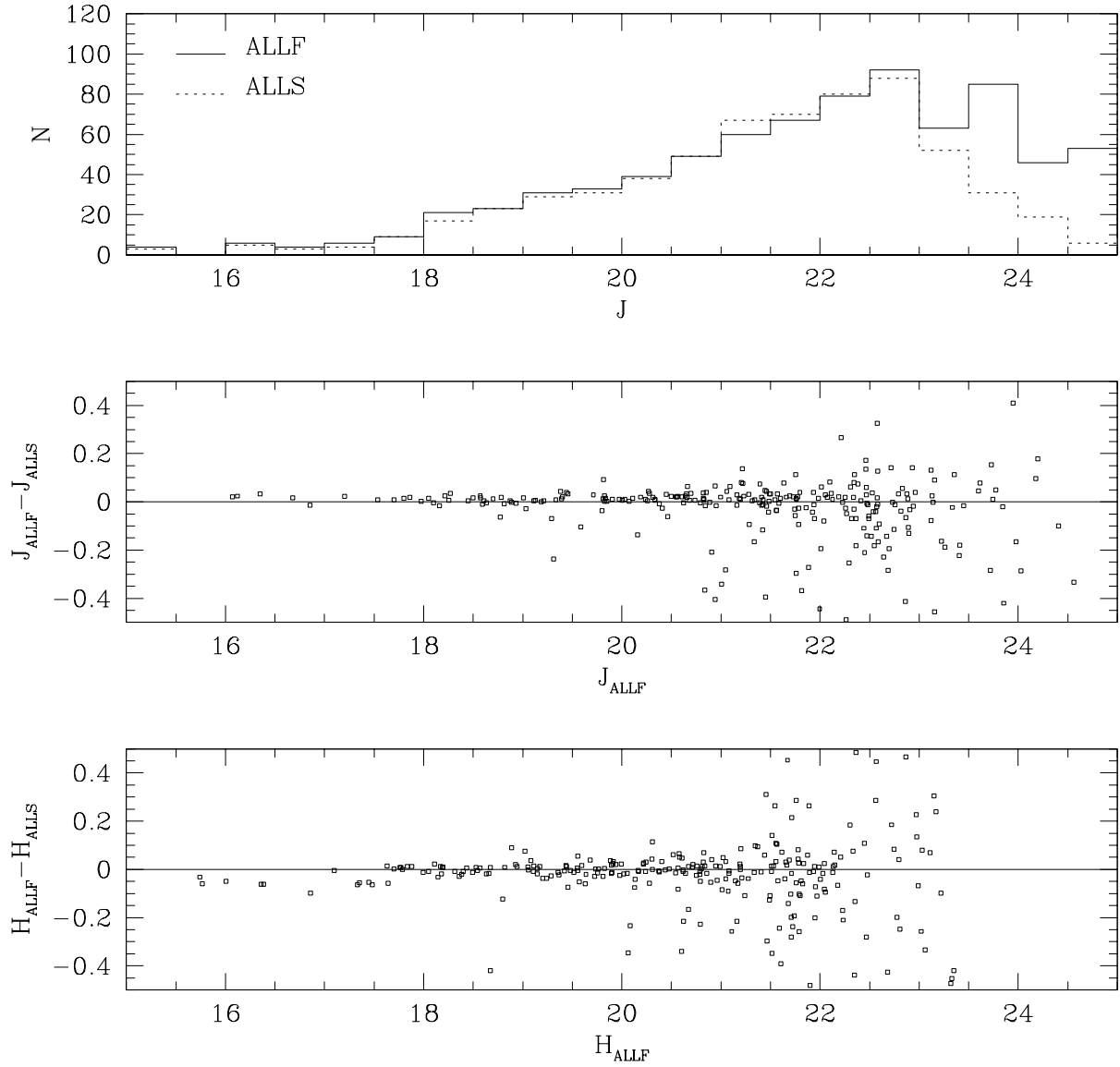,width=17truecm}}
\caption[]{Comparison between the results of ALLFRAME, as 
described in \S\ \ref{phot} (index ALLF) and the results of the 
standard DAOPHOTII/ALLSTAR procedure (index ALLS).}
\label{andy}
\end{figure}

A second, independent reduction of the data was carried out 
with the same software (DAOPHOTII/ALLSTAR) but with somewhat different
procedures.  Nearly identical results were obtained as in the first
reduction.  In this reduction we performed simple
star finding and PSF fitting on each individual frame but without
using the median image, or ALLFRAME.  The resulting photometry is
somewhat shallower but it provides a useful consistency check both in
terms of magnitudes and numbers of identified stars, in the common
magnitude range. Figure~\ref{andy} shows the comparison between the
(calibrated) output photometry of the two procedures.  The two bottom
panels show that, for the stars identified in both cases (i.e., except
for the fainter ones, identified only by ALLFRAME) the measured
magnitudes are in very good agreement, with a very small offset
$\Delta J=0.01$, due to some systematic error in one (or both) of the
aperture corrections. The $H$ magnitudes of the brightest stars
($H<17.6$) also differ by $\Delta H\sim0.05$.  The top panel shows the
two LFs (before completeness correction) which are almost identical down to
$J=23$, where the ALLFRAME reductions go significantly deeper.

\section{The Color--Magnitude Diagram}
\label{seccmd}

\begin{figure}
\vskip 0.4cm
\centerline{\psfig{file=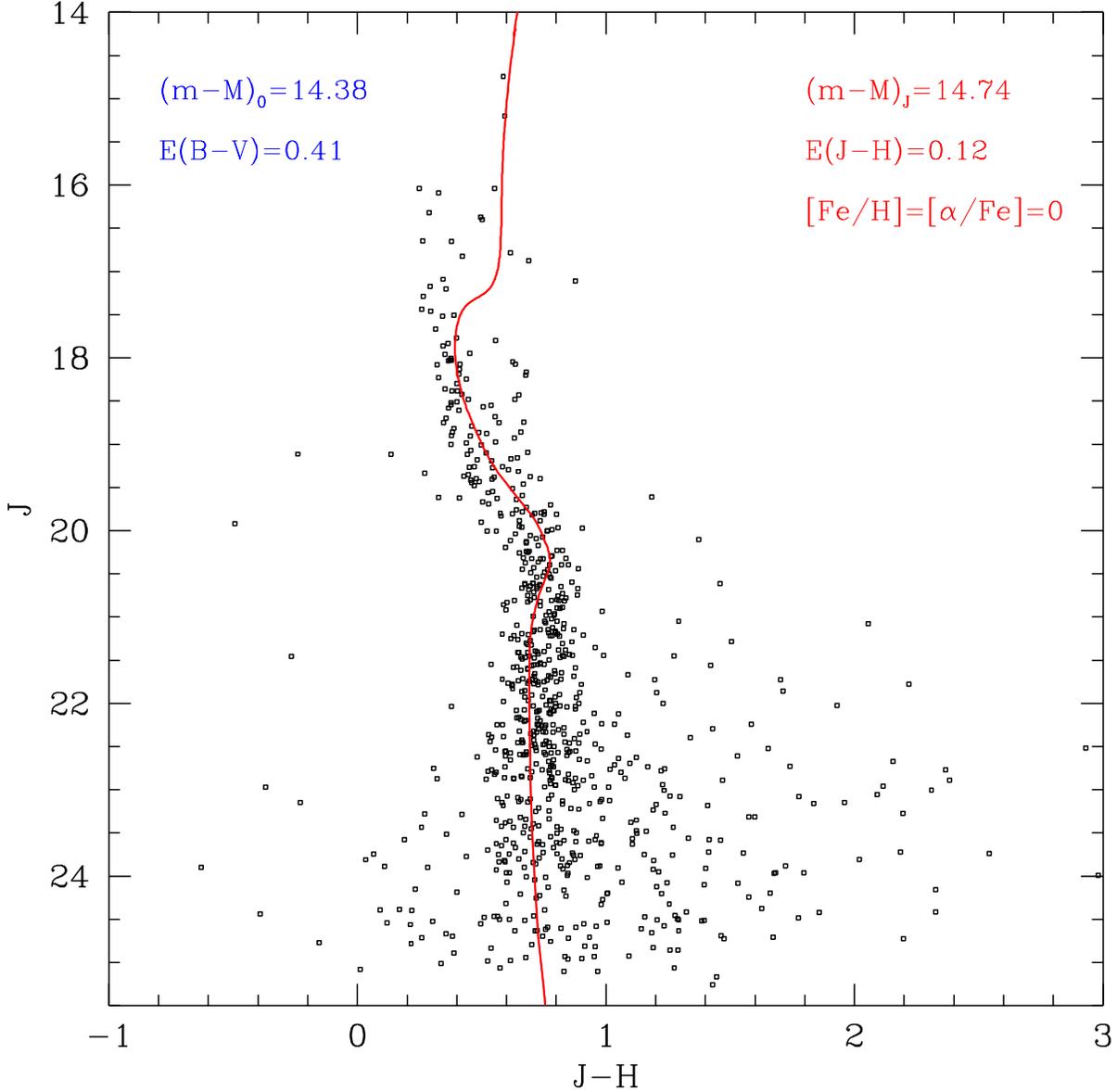,width=17truecm}}
\caption[]{The observed color--magnitude diagram of the 780 stars in 
the NIC2 field. The solid line is the isochrone for a solar
metallicity, 10 Gyr old population (Cassisi et al. 1999). The labels
on the left indicate the values of absolute distance modulus and
reddening taken from the literature (Rich et al. 1998), while the
values on the right quote the corresponding quantities in the infrared
bands, adopted for this comparison.  The shift to redder colors of the
isochrone at $J\sim20.5$ ($M_{\rm J}\sim5.75$) is caused by the appearance of 
opacity from water and molecular hydrogen.}
\label{cmd_iso}
\end{figure}

The observed color--magnitude diagram (CMD) for the 780 stars
measured in our $22\secondip 5 \times 22\secondip 5$ field is shown in
Figure~\ref{cmd_iso}. Only the stars identified in at least 5 independent
frames per filter are plotted.  A further selection on the magnitude
error and on the {\it sharp} parameter was imposed to
discard spurious detections due to noise and intersecting diffraction
spikes that may remain around the brightest stars.  The bulge main
sequence (MS) is well defined from the turnoff ($J\sim18$) down to
magnitude $J\sim24$ where the sequence starts to broaden and the
density of stars falls abruptly.  A prominent feature in this CMD is
the sharp bend that is clearly visible at $J\approx20.5$. Fainter than
this point, the MS is almost vertical.  As predicted by stellar models
(Cassisi et al. 1999; Baraffe et al. 1997), this behavior is due to
the competition between the tendency towards redder colors due to both
the decreasing effective temperature and the increasing molecular
absorption at optical wavelengths, and the increasing
collision-induced absorption of molecular hydrogen at infrared
wavelengths (CIA mechanism, Saumon et al. 1994).

The two brightest stars in the CMD of Figure~\ref{cmd_iso}, located in the
left side of our field (Fig.~\ref{field}), were saturated; their
magnitudes have been measured independently by extrapolating their PSF
profiles into the central region.

Also shown in Figure~\ref{cmd_iso} is the theoretical isochrone by
Cassisi et al. (1999).  These models have been constructed by adopting
the most updated input physics, such as stellar opacities, equation of
state, and outer boundary conditions (see Cassisi et al. 1999, for
more details). We adopt here the absolute distance modulus and
reddening of this region of the Galactic bulge, as measured by Rich et
al.\ (1998): $(m-M)_0=14.38$ and $E(B-V)=0.41$. By assuming $R_V=3.1$
the extinction is $A_V=1.27$, which can be converted to the
corresponding $A_J$ and $A_H$ by means of the relations given by
Cardelli et al.  (1989): $A_J=0.282A_V$ and $A_H=0.190A_V$. The
isochrone shown in Figure~\ref{cmd_iso} refers to solar metallicity
([Fe/H]=[$\alpha$/Fe]=0) and an age of 10 Gyrs. The model is a
satisfactory match to the general shape of the observed MS, in
particular the position of the bend at $J\sim20.5$ ($M_{\rm
J}\sim5.75$) is well reproduced, even if its strength seems to be a
little overestimated.  This feature also provides a good check of the
zero point of the photometric calibration and the adopted distance and
reddening.

The present NICMOS data provide a too sparse sampling of the turnoff area
for properly addressing the issue of the age of the bulge stellar populations. 
This will be attempted in a future paper, combining our NICMOS data with 
deep WFPC2 observations of the same field, as well as wide field $V$ and $I$
observations taken at the ESO/MPIA 2.2m telescope (Zoccali et al. 1999).


\section{The Luminosity Function}
\label{seclf}

In order to obtain the stellar LF of our field, particular attention
was devoted to estimating the completeness of our sample. Standard
artificial-star tests were carried out on the NIC2 field, in the same
way as described in detail by Piotto \& Zoccali (1999).  We performed
ten independent tests, by adding about seventy stars each time, with
magnitudes in the range $20<J<25$. Visual inspection of the
stars-subtracted image insured that our photometry was complete for
brighter magnitudes.  The artificial stars were arranged in a spatial
grid such that the separation between the centers of each star pair
was two PSF radii plus one pixel. This allowed us to add the maximum
number of stars, without creating overcrowding.  In addition, the
position of each star in the grid was randomly located inside one
pixel, so as to prevent the
centers of all the artificial stars from falling on the same position
within a pixel, which would have biased their probability of being
detected. The artificial stars were added on each individual $J$ and $H$
image. It should be noticed that the stars must be added in the same
position {\it on the sky}, therefore their coordinates must be
different in different frames, following the frame-to-frame
coordinate transformations calculated from the original photometry. A
high precision is required in this process, in order to be able to
measure the artificial stars with the same photometric accuracy of the
original ones. We then ran the same photometry procedure used for the
original photometry: star finding was performed on the median of all
the star--added images, and then ALLFRAME was used for the
simultaneous photometry of all the frames. The same selection criteria
used for the original stars were applied to the output list of the
artificial star tests.

\begin{figure}
\vskip 0.4cm
\centerline{\psfig{file=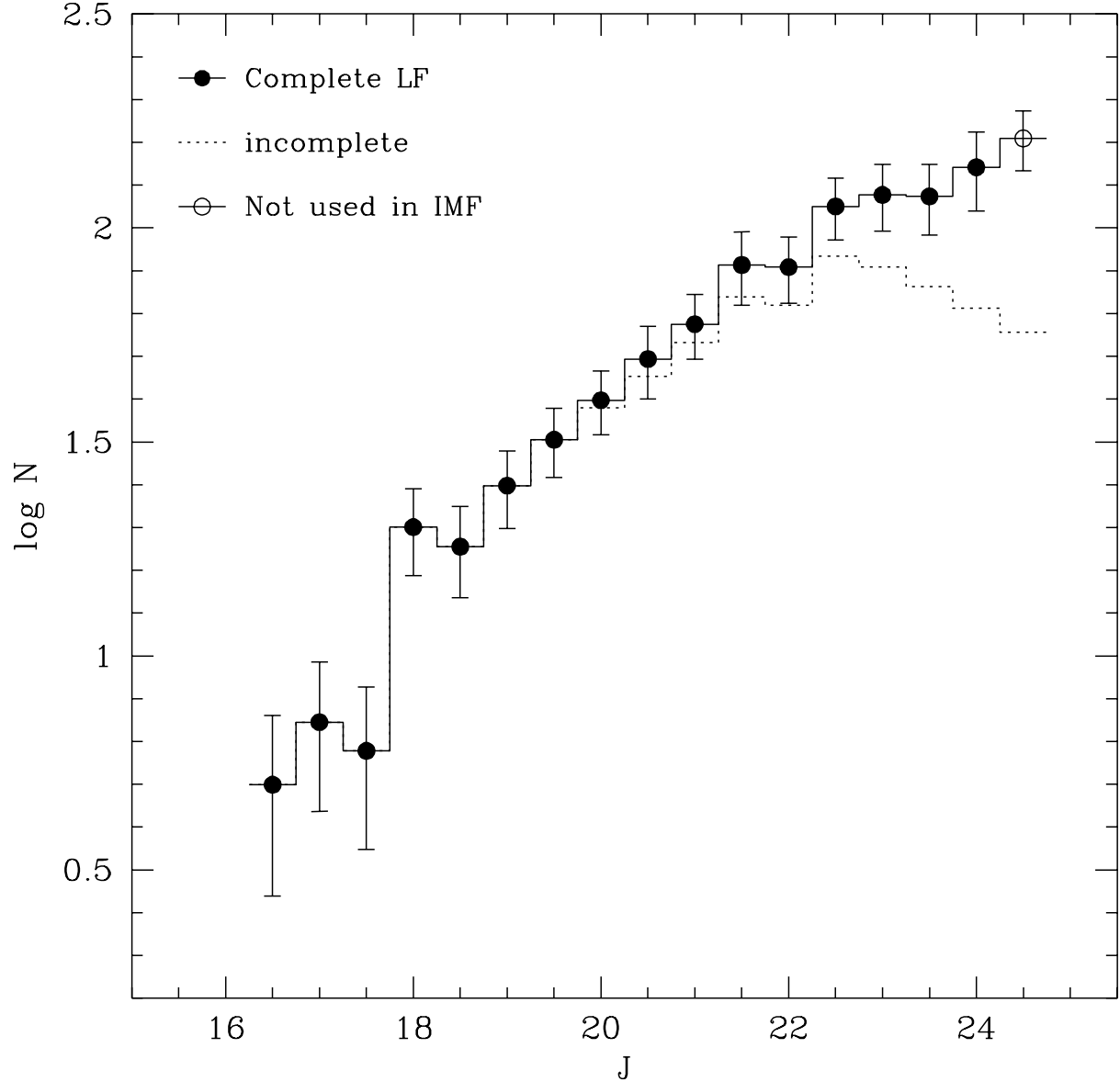,width=17truecm}}
\caption[]{The LF extracted from the CMD of Fig.~\ref{cmd_iso}. The
LF before the completeness correction is shown as a dotted
histogram. Error bars are the quadratic sum of the Poisson error on
the raw star counts and the error in the completeness corrections.
The faintest bin was not used in the derivation of the IMF, because of its 
low completeness.}
\label{lf}
\end{figure}

The completeness correction obtained in this way was applied to the LF
obtained from the CMD of Figure~\ref{cmd_iso}.  This procedure also
automatically compensates for the differences of the total integration
time across the $22\secondip 5\times 22\secondip 5$ field.  It is
worth noting that the scatter in the color of the stars on the right
of the main sequence, for $J>21$, is also present in the CMD for the
artificial stars, which indicates that the effect is spurious.  Visual
inspection of these stars on the image revealed that they are all
located on the left side of the field, where scattered light from a
few very bright objects is also present. Some of them could be
residual noise spikes, but some are likely to be real stars whose
magnitude has been enhanced due to the proximity of brighter stars.
The fact that these objects are present only on the right side of the
main sequence indicates that such effect is stronger for the $H$
magnitudes, a likely result of the poorer PSF in the $H$-band.  
The way in which we applied the completeness correction
(i.e. determining the completeness fraction as a function of
the recovered magnitude of the artificial stars, instead of the input
magnitude) automatically takes into account the effect of the migration
of the stars towards brighter magnitudes, therefore we didn't impose
any further selection on the CMD of Figure~\ref{cmd_iso}.  The
resulting $J$-band LF is shown in Figure~\ref{lf}; it is very smooth over
the whole range from the bin at $J=18$ (turnoff region) to the faint
limit at $J=24$. Also shown as a dotted histogram is the raw LF,
without completeness correction. In the determination of the IMF, we
did not use the first two bins that, according to our model,
correspond to evolved stars, nor the very last bin ($J=24.25$)
because its completeness is $\sim30\%$. The second to last bin, at
$J=24$, is complete at $46\%$.


Since the field is located at low Galactic latitude, contamination by
disk stars cannot be neglected.  We offer an estimate of this
contamination using Kent's (1992) model for the $K$ band luminosity
distributions of the disk and bulge. If the LFs of the disk and bulge
have the same form as the observed LF in our field, scaled for
distance and stellar density, then we find that about 11\% of the
stars in our field are disk stars, with a modest trend from about 9\%
at the faint end ($J\sim 24$), to about 14\% 
at the bright end ($J\sim 17$)
of the LF in Figure~\ref{lf_fro}.  We then adopt an overall reduction
of the LF by 11\% for $J>17$.  We note, however, that this is only a
rough estimate of the disk contamination.  Available data
do not allow a more accurate correction, and further optical
and IR data would be required to address this problem more properly.
The LF of Figure~\ref{lf} is listed in Table~\ref{tab_lf}. Column 1
gives the $J$ magnitude, Column 2 and 3 give the raw and completeness
corrected counts, respectively, Column 4 is the error, and Column 5 the
estimated contribution from disk stars.

Contamination by extragalactic objects is estimated using the NICMOS
$H$-band galaxy counts (Yan et al. 1998). This gives less than one
galaxy for $J<23$, and between 1.2 and 2.9 galaxies in the last two
bins of our LF, corresponding to $J=23.5$ and $J=24$, respectively. We
therefore conclude that this source of contamination can be neglected
in our analysis.


\section{NIC1 data}
\label{secnic1}

\begin{figure}
\vskip 0.4cm
\centerline{\psfig{file=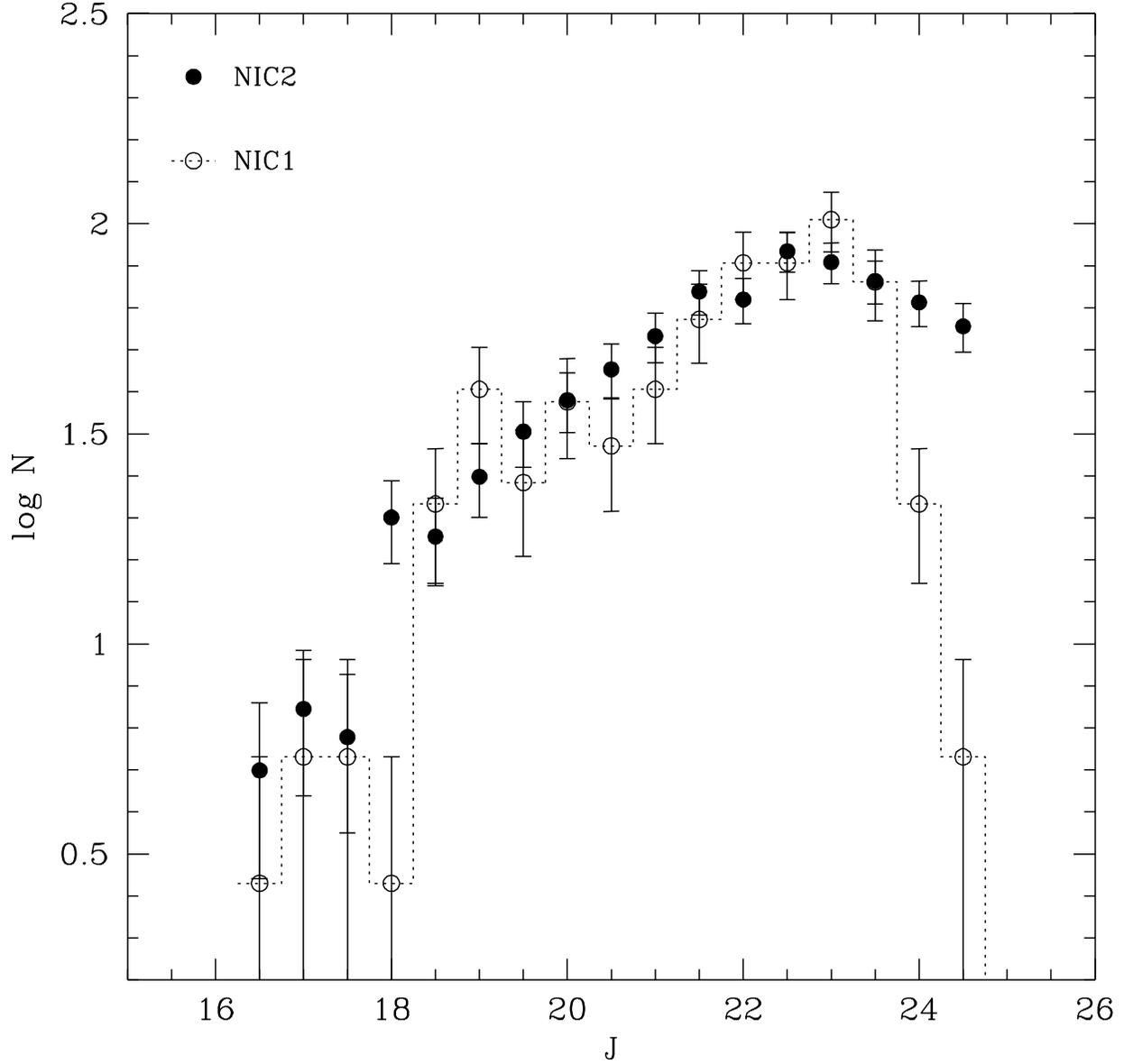,width=17truecm}}
\caption[]{Comparison between the LF extracted from the NIC2 and
NIC1 data.  Neither data set is corrected for incompleteness. The two
LFs are very similar, down to magnitude $J\approx23$ where the NIC1 LF
falls abruptly due primarily to the {\it by eye} selection adopted to
eliminate spurious detections.}
\label{lfnic1}
\end{figure}

An additional set of data on a nearby bulge field is provided by our
parallel observations with the NIC1 camera through the F110W filter.
The field of view of NIC1 is significantly smaller
($11^{''}\times11^{''}$) than that of NIC2, but thanks to its smaller
pixel size (0$\secondip$043), it allows more accurate sampling of the
PSF and therefore it yields more accurate photometry for stars with
good photon statistics. In our case, due to the rotation of the NIC2
field in different visits, the NIC1 camera actually mapped a larger
region ($16^{''}\times14^{''}$). The use of only one filter has the
disadvantage that it is not possible to construct a CMD, but allowed a
longer exposure time (35850 s).

We expect NIC1 to be more complete than NIC2 at intermediate
magnitudes because of its better resolution and longer exposure time
(and hence higher S/N).  At faint magnitudes the higher NIC1 read
noise implies that NIC1 data should have only slightly better S/N despite
the longer exposures.  Nevertheless, we expected to be able to push
the LF to somewhat fainter magnitudes by incorporating the NIC1 data.
We reduced the NIC1 frames with the same algorithm adopted for NIC2,
and were able to measure about 800 stars. Unfortunately, the selection
criteria adopted for NIC2 were not sufficient to ensure ``clean''
photometry in this case because in all the NIC1 frames there was a
shaded region, apparently due to some flat-field problems. Many faint,
possibly spurious stars were identified in this region, and we were
not able to find a suitable selection criterion to discard them
without also losing what seemed to be real stars.  Due to this
problem, and since we had no CMD to guide the selection between real
stars and spurious detections, we decided to check each of the 800
stars by eye on the image. This certainly introduced a brighter
magnitude limit, because it was hard to make a selection in the very
last magnitude bin, and also prevented us from making any artificial
star test, as the selection criterion was not automatic.  Thus,
despite our expectations, the LF extracted from the NIC1 photometry
could not be used to extend the NIC2 LF to fainter magnitudes.
However, it is useful as a cross check on the NIC2 results.
Figure~\ref{lfnic1} shows the LFs extracted from the NIC1 and NIC2 data,
with no correction for incompleteness.  The
two LFs were normalized according to the relative areas of the two fields.
The LF from NIC1 falls abruptly below $J=23$, due
primarily to the visual selection to eliminate spurious
detections. However, it is reassuring to note that the two LFs track
each other very closely, down to magnitude $J=23$.


\section{The Mass Function}
\label{secmf}

\begin{figure}
\vskip 0.4cm
\centerline{\psfig{file=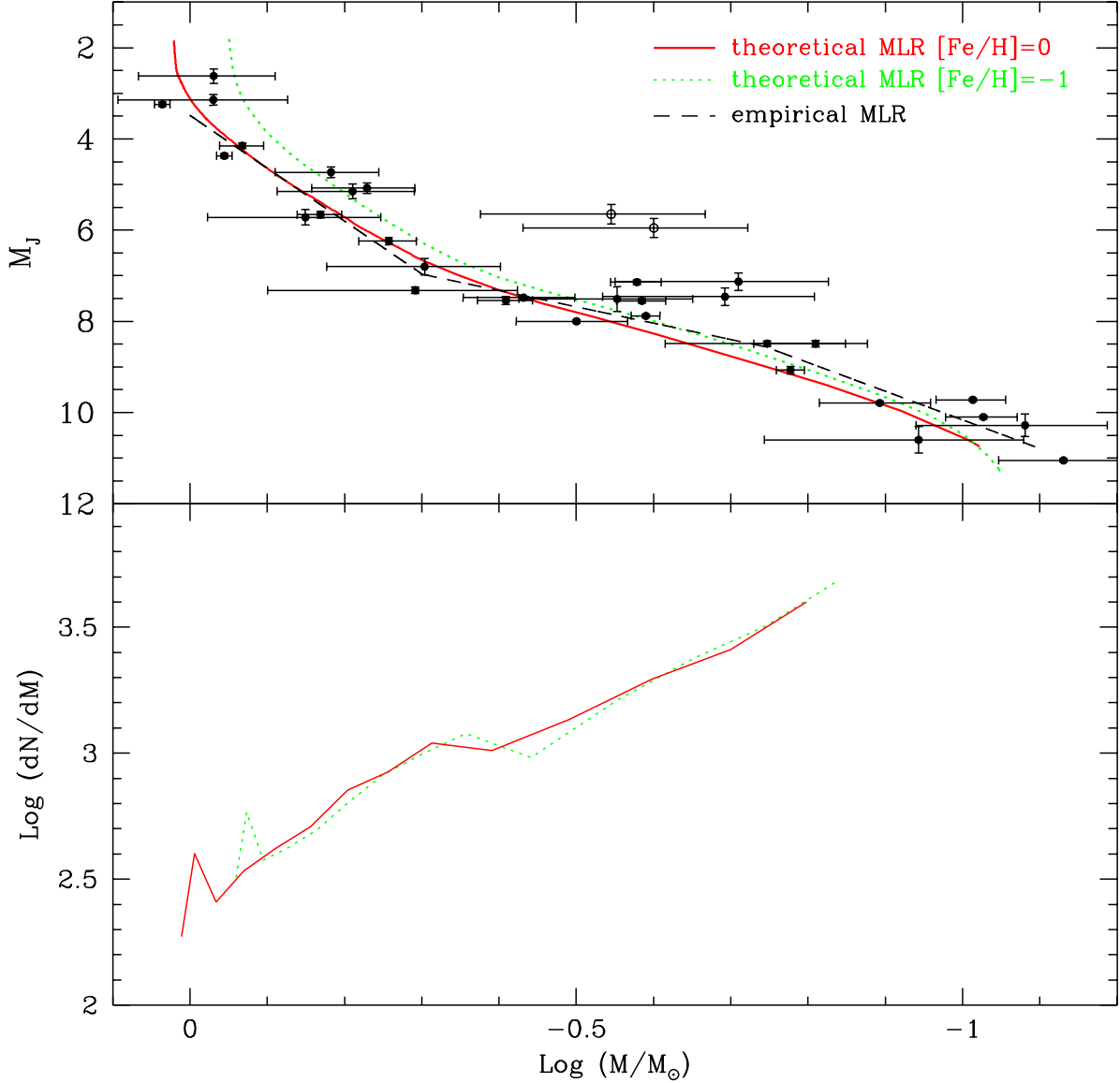,width=17truecm}}
\caption[]{Top panel: The adopted theoretical MLR (solid line), 
for solar metallicity (Cassisi et al. 1999), is compared with the
measured masses and luminosities for solar neighborhood stars (Henry \&
McCarthy 1993) (dots) and with the empirical MLR (dashed line)
suggested by the same authors. Also shown as a dotted line is the
theoretical MLR for a metallicity of [M/H]=--1.0 to emphasize that, in
the infrared bands, even such a large change in metallicity would not
significantly change the slope of the MLR. In the bottom panel the IMF
obtained from the adopted model MLR ([M/H]=0) is compared with the one
we would obtain using the MLR for [M/H]=--1.}
\label{mlr}
\end{figure}

The LF for low-mass stars ($M\lsim1M_\odot$) can be converted into a
MF which is the same as the IMF, since the stars are unevolved and
their number is unaffected by dynamical processes.  In order to
transform the LF into the IMF a mass--luminosity relation (MLR) is
required. An empirical MLR in the infrared bands has been determined
for solar metallicity stars by Henry \& McCarthy (1993) from a sample
of visual and eclipsing binaries in the solar neighborhood. This MLR
is shown as a dashed line in Figure~\ref{mlr} (top panel) together
with the individual data points.  This relation was obtained from a
series of quadratic fits in different mass intervals, and would
introduce features in the IMF at each of the abrupt changes in the
slope of the MLR. Also shown in Figure~\ref{mlr} are the MLRs for two
sets of theoretical models (Cassisi et al.  1999).  The empirical and
theoretical MLRs are in very good agreement, apart from a discrepancy
by few hundredths of a solar mass near the faint limit. Given the
large spread and errorbars of the data points, this small discrepancy
appears to be completely negligible. From this comparison, and the
good fit of the CMD of Figure~\ref{cmd_iso}, we feel confident in
adopting the theoretical MLR to convert the observed LF into an IMF.
The MLR for solar metallicity is adopted in the present work.
However, as shown in Figure~\ref{mlr} the metallicity dependence of
the MLR is so small that we would expect no appreciable effect even if
the average metallicity of the stars in our bulge sample were very
different from solar, which is not the case (McWilliam \& Rich
1994). This is illustrated in the bottom panel of Figure~\ref{mlr},
where the IMF obtained from the adopted model MLR ([M/H]=0) is
compared with the one obtained using the MLR for [M/H]$=-1$.

The resulting IMF for the Galactic bulge is shown in
Figure~\ref{mf_bulge}.  Within the errors, the IMF can be represented
over the entire mass range by a single power-law of the form
$dN\propto M^\alpha dM$, having a slope $\alpha=-1.33\pm0.07$ (where
Salpeter has $\alpha=-2.35$).  It is worth noting that adopting the
widely used distance modulus $(m-M)_0=14.5$ to the Galactic center
(Reid 1993), instead of the value of 14.38 adopted for this paper, the
resulting slope is $\alpha=-1.30$ over the whole range
$1<M/M_\odot<0.15$.  Hence the result is fairly insensitive to errors
in the distance.
As it appears from Figure~\ref{mf_bulge}, there is a hint for the IMF
to flatten slightly below $0.5M_\odot$. A two-slope IMF gives indeed
a marginally better fit, with $\alpha=-2.00\pm0.23$ for $M>0.5M_\odot$,
and $\alpha=-1.43\pm0.13$ for $M<q0.5M_\odot$.

To some extent the presence of binaries can introduce a bias in the
derived IMF slope.  However, the frequency of binaries in the bulge
and the distribution of their mass ratios remains unconstrained by the
present data, and therefore we do not simulate the effect of binaries
in this work.  Holtzman et al. (1998) assumed various binary fractions
(defined as the fraction of systems that are binaries) up to 50\%.
They find that the slope at the faint end steepens by 0.4 for a binary
fraction of 50\% (2/3 of all stars in a binary system, with both
primaries and secondaries following the same IMF). We conclude that
adopting the same procedure of Holtzman et al. would bring $\alpha$
from $\sim -1.3$ to $\sim -1.7$ for such an extreme fraction of
binaries.


\section{Comparison with other mass functions}
\label{mfs}

\begin{figure}
\vskip 0.4cm
\centerline{\psfig{file=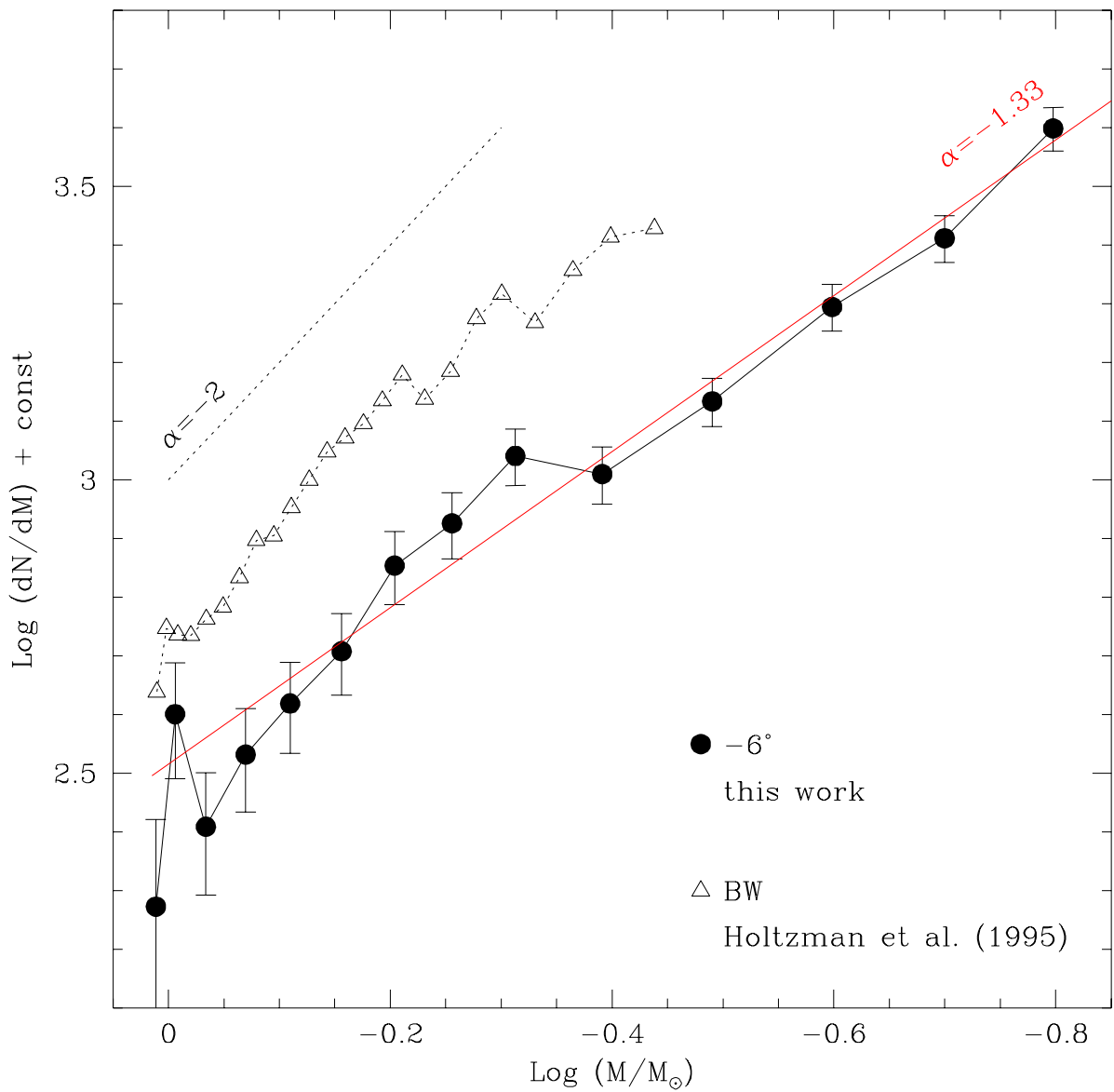,width=17truecm}}
\caption[]{The IMF for the $-6^\circ$ field (solid symbols). 
A single power-law with a slope of $\alpha=-1.33\pm0.07$ is able to
fit the data in the whole mass range.  Were the fit restricted to
$M>0.5M_\odot$, a steeper slope $\alpha=-2.0\pm0.2$ would be obtained
(dotted line). The quoted errors on the slopes are the formal errors
on the fit. Also shown is the Baade's Window IMF from Holtzman et
al. (1998)}
\label{mf_bulge}
\end{figure}

\begin{figure}
\vskip 0.4cm
\centerline{\psfig{file=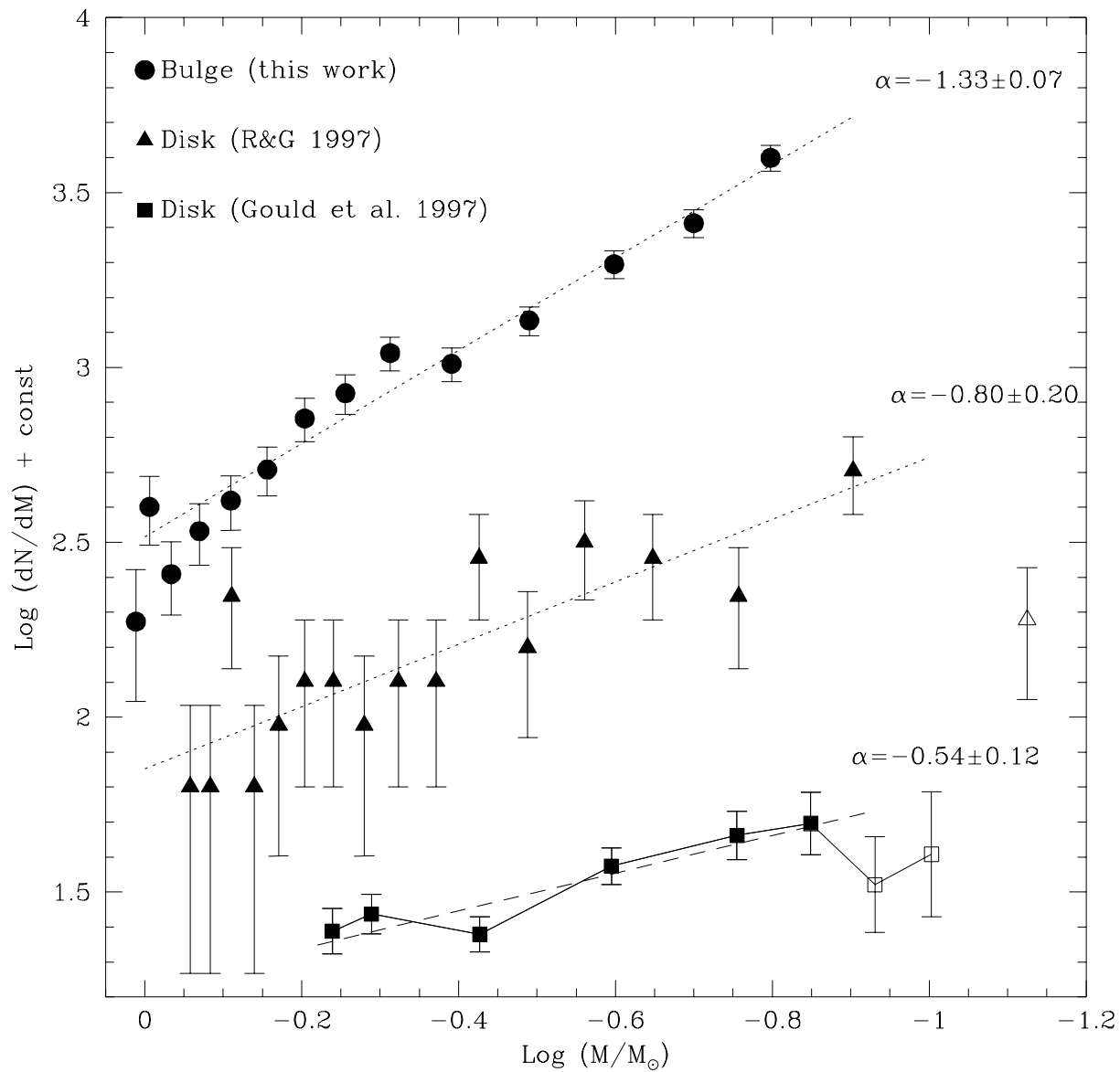,width=17truecm}}
\caption[]{The bulge IMF shown in Fig.~\ref{mf_bulge} is compared with two
independent determinations of the IMF of the Galactic disk.  The
values shown in the figure, $\alpha=-0.8$ for Reid \& Gizis (1997) and
$\alpha=-0.54$ for Gould et al.\ (1997) are based on our fits to the
data, based on the restricted range $M>0.15\,M_\odot$ (solid
symbols) and therefore differ slightly from the values reported by
the original authors.}

\label{mf_disk}
\end{figure}

Figure~\ref{mf_bulge} shows the comparison between the IMF derived
above and the IMF from Holtzman et al. (1998) for the stars in Baade's
Window. For a more consistent comparison, we used the $I$-band LF from
their Figure~\ref{lf_fro}, and converted it into an IMF by means of
the same theoretical MLR that we have used for our derivation (except
for the color transformation to the $I$ band instead of the $J$ band),
and without applying a correction for binaries.  The derived IMF is
very similar to the one that Holtzman et al. (1998) originally
obtained down to $\sim 0.35$ with an empirical MLR which is in
good agreement with that of Henry \& McCarthy (1993). The IMF so
derived turns out to be very similar to our bulge IMF for the
$-6^\circ$ field. Holtzman et al.'s Baade's Window IMF has a slope
$\alpha=-2.2\pm0.2$ for $M>0.5M_\odot$, and $\alpha=-1.4\pm0.2$
for $M<0.5M_\odot$, virtually identical to our result over the wider
mass range that was accessible for the $-6^\circ$ field.

It is also interesting to compare the bulge IMF with the IMF of the
Galactic disk. The comparison is shown in Figure~\ref{mf_disk}, which
displays our bulge IMF together with the disk IMF as recently derived
by Reid \& Gizis (1997) and by Gould et al. (1997).  Reid \& Gizis
(1997) extract their LF from the study of a volume-complete sample of
low mass stars with $\delta>-30^\circ$ and within 8 pc from the Sun.
Their IMF, shown in Figure~\ref{mf_disk}, is not corrected for binary
stars, and therefore it can be compared with the IMF we derive for the
Galactic bulge.  Note that this disk IMF has also been derived using
the Henry \& McCarthy (1993) empirical MLR. The Reid \& Gizis (1997) 
disk IMF is well represented by a power law, but its slope 
$\alpha=-0.80\pm0.20$ differs by about $2.4\,\sigma$ from that of 
the bulge.  The disk IMF by Gould et al. (1997) is also shown in 
Figure~\ref{mf_disk}.  For their sample of disk M dwarfs they found an 
IMF with a slope $\alpha=-0.54\pm0.12$, in $1\,\sigma$ agreement with 
Reid \& Gizis (1997) disk IMF, but definitely flatter than the bulge IMF.
On the contrary, the IMF for the $-6^\circ$ field is in very good
agreement with the disk IMF obtained by Kroupa et al. (1993), for stars
within $\sim 5$ pc from the Sun, which has a slope $\alpha=-2.2$ for 
$0.5<M/M_\odot<1$ and $\alpha=-1.3$ for $0.08<M/M_\odot<0.5$.

Similar results come from the comparison of the bulge IMF with the MF
of young open clusters. Recent work in this field has been done by
Hillenbrand (1997), Luhman et al. (1998) and Bouvier et al. (1998).
From an extensive optical study of the Orion Nebula Cluster,
Hillenbrand (1997) found an IMF slope of $\alpha\approx-1.35$ for
$0.2<M/M_\odot<1$, but also a sharp peak at $0.2 M_\odot$ and a
turnover for lower masses. In contrast, both Luhman et al. (1998),
for the young cluster IC 348, and Bouvier et al. (1998), for the 
Pleiades, found a log-normal IMF that matches that of Miller \&
Scalo (1979) for $M\lsim0.3$ and a flatter IMF, with slope
$\alpha=-0.6$, for lower masses.

\begin{figure}
\vskip 0.4cm
\centerline{\psfig{file=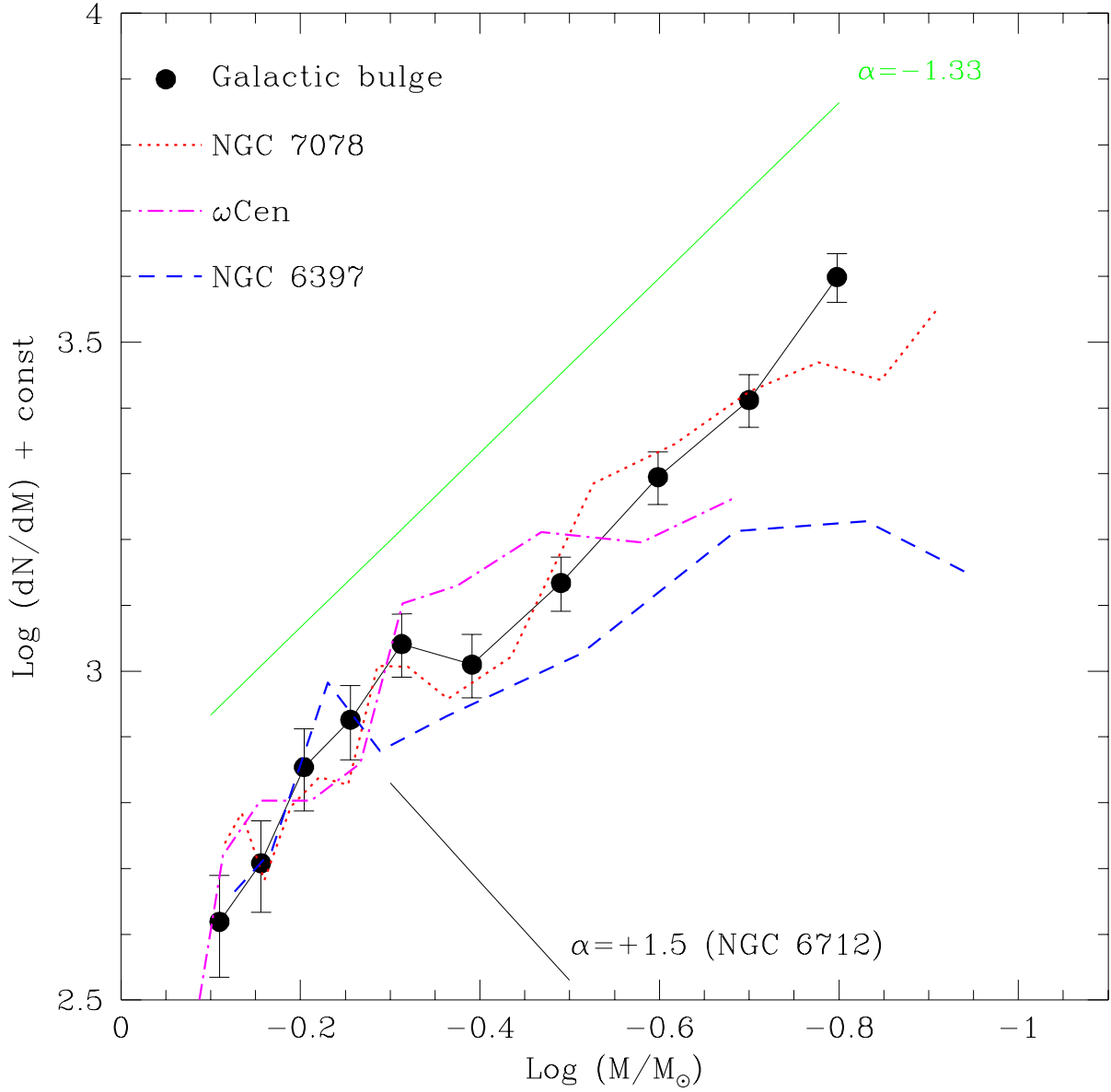,width=17truecm}}
\caption[]{The bulge IMF is compared with the MFs observed in
a sample of Galactic globular clusters.}
\label{ggc}
\end{figure}

Finally, it is interesting to compare the bulge IMF with the MF
measured in some Galactic globular clusters (GCs). GCs
are strongly affected by dynamical evolution which modifies their stellar
MF.  Several GCs have short relaxation times with respect to
their age, and therefore their observed MF changes with radius due to
mass segregation and evaporation.  They are also affected by tidal
shocks caused by passage through the Galactic disk and bulge that
preferentially strip lower mass stars.
According to dynamical models (Vesperini \& Heggie 1997) the only way
to measure a MF unaffected by these dynamical processes is to observe
GCs with high mass (i.e., long relaxation time) and very wide orbits
(i.e., that do not cross the Galactic plane frequently).  
In Figure~\ref{ggc} the IMF of
the Galactic bulge is compared with the MF of a few GCs. Clusters with
extreme MFs were chosen to make the clearest plot. The MF of NGC~7078
(Piotto et al. 1997), a massive ($\log M$=6.3, Djorgovski 1993) very
metal poor cluster with very wide orbit (Dauphole et al. 1996), is
very similar to the IMF of the Galactic bulge. The MF of $\omega$ Cen
(Pulone et al. 1998), very massive ($\log M$=6.6) but with a smaller orbit,
is only slightly flatter, while the MF of NGC~6397 (King et al. 1998)
is significantly flatter, this cluster having a smaller mass ($\log
M=5.4$) and a very tight orbit.  Most extreme is the case of NGC~6712
(De Marchi et al. 1999) that has a MF with an inverted slope
$\alpha=+1.5$. This is a low mass cluster ($\log M=5.0$), with an
orbit that brings it to within $\simeq 300$ pc of the Galactic center.
Our finding that the bulge IMF is similar to that of the less
dynamically affected clusters suggests that the GCs and the bulge may
have the same IMF.  The similarity of the IMF of the solar metallicity
bulge with that of NGC 7078 at $\rm [Fe/H]=-2$ suggests that the slope
of the IMF is relatively independent of metallicity (see also
Grillmair et al. 1998).


\section{Implications}
\label{disc}

In this section we briefly discuss a few implications and applications
of the bulge IMF derived in the previous sections.


\subsection{A Complete Bulge Luminosity Function}
\label{lfbright}

The bulge LF extending from near the MS turnoff down to the lower MS
can be combined with an appropriate LF for bright, evolved stars in
the bulge.  This approach permits us to construct a complete bulge LF
that extends from the tip of the red giant branch (RGB) to nearly the
bottom of the MS, which can be compared with theoretical stellar
evolution models, and which can be used as a template in a variety of
applications.  To this end, we have combined our LF with the LF of
Tiede et al. (1995), appropriately scaled by its 8.01 times greater
area (4056 arcsec$^2$ versus 506 arcsec$^2$).  The Tiede et al. field
is located only 10 arcmin from our bulge field, and thus should have
essentially the same stellar population.  The brightest part of the LF
$(J<11.5)$ is adapted from the wide-area survey of bulge M giants in
Baade's Window $(b=-4^\circ)$ (Frogel \& Whitford 1987), properly
normalized to the 506 arcsec$^2$ NIC2 field both for the area, and
for the lower surface brightness of our field.

\begin{figure}
\vskip 0.4cm
\centerline{\psfig{file=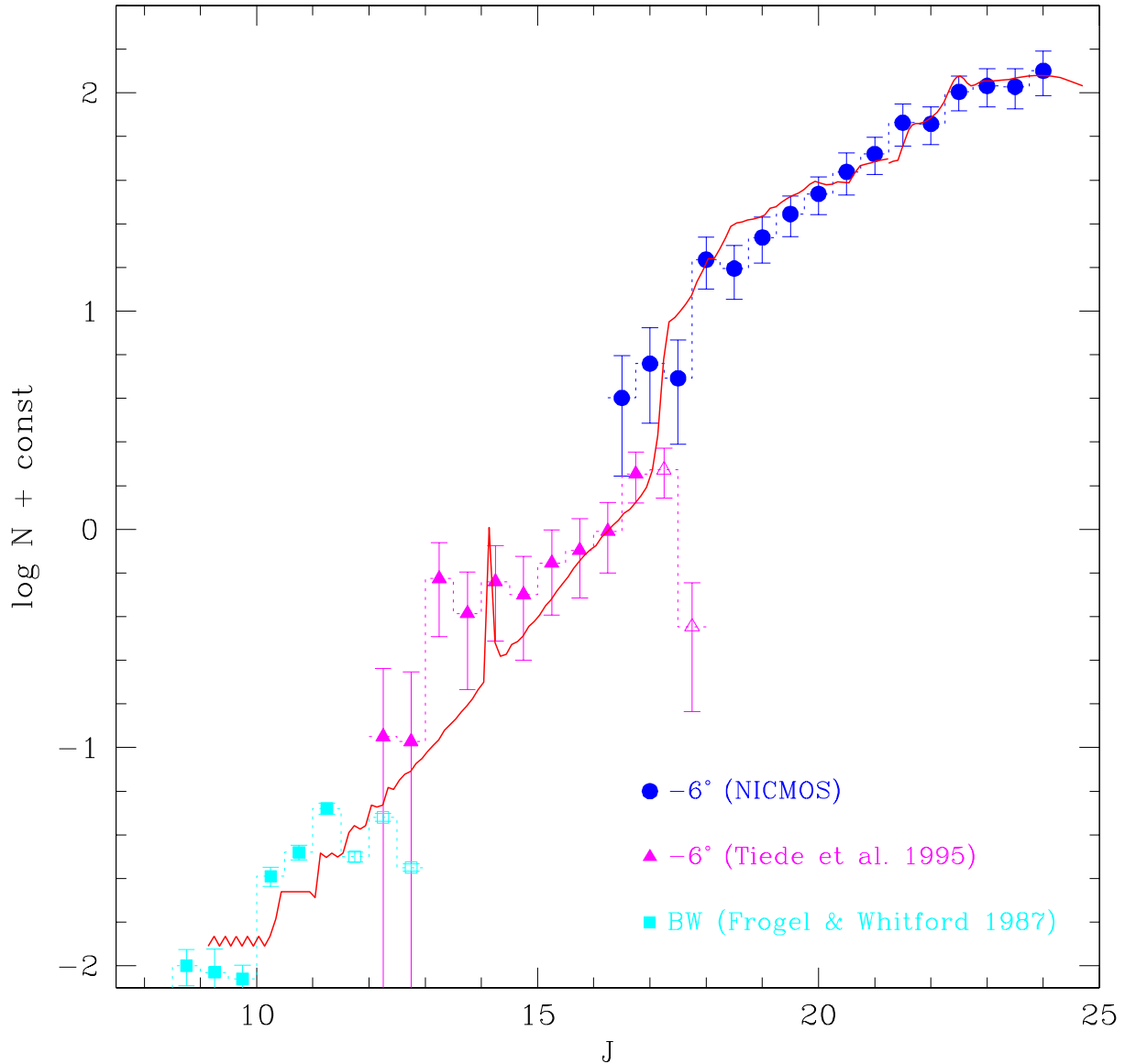,width=17truecm}}
\caption[]{The LF shown in Fig.~\ref{lf} is extended to bright magnitudes 
using the LF from Tiede et al. (1995).  The brightest stars (filled
squares) are taken from the wide-field survey of M giants in Baade's
Window (Frogel \& Whitford 1987). Filled symbols show complete
counts, while open symbols refer to counts affected by but not
corrected for incompleteness. The solid line is the theoretical LF for
a 10 Gyr, solar metallicity population, with IMF slope $\alpha=-1.33$
(see text).  Note that the observed peak at $J\sim 13.5$ is the red
clump of the HB stars, not included in the models, while the sharp
peak in the theoretical LF at $J=14.15$ is the RGB bump.}
\label{lf_fro}
\end{figure}

These LFs have been corrected for the disk contamination. The disk
contribution to the MS was evaluated as described in \S\ 4. For the
stars brighter than the MS turnoff the fraction of disk stars was
estimated in a more direct way.  We used $V$ and $I$ observations of a
wide region including the small NIC2 field, taken with the 2.2m
telescope + the Wide Field Imager (WFI) at ESO La Silla, on 24 March
1999 (Zoccali et al. 1999).  The CMD derived from these images is very
well populated from the tip of the RGB down to about 2 magnitudes
below the turnoff, and allows one to separate very clearly the
extended disk MS from the evolved population of the bulge: RGB +
horizontal branch (HB) + asymptotic giant branch (AGB). Lines of
constant $J$ magnitude drawn on the $(V,V-I)$ CMD using color
transformations (Allard et al. 1999) allow one to count the number of
disk and bulge stars in each $J$ bin. The decontaminated LF,
renormalized to our NIC2 field is reported in Table~\ref{tab_disk}
along with the value of the decontamination correction. Therefore, the
LFs in Table~\ref{tab_disk} and Table~\ref{tab_lf} have the same
normalization, and the resulting global LF is shown in
Figure~\ref{lf_fro}.  Superimposed on this empirical LF is the
theoretical LF from models by Cassisi et al.  (1999), extended to the
tip of the RGB using models by Bono et al. (1997). Note that the
theoretical LF does not include either the HB clump (clearly visible
in the empirical LF) or the AGB.  The sharp peak at $J=14.15$ is the
RGB bump, which is produced by the pause in evolution along the
RGB when the hydrogen shell burns through the hydrogen discontinuity
left by the deepest penetration of the convective envelope.  This
feature is not very clear in the observed LF because it is smeared by
distance dispersion and differential reddening, and partially merged
with the HB red clump. Both the slope of the RGB LF and the sharp drop
between MS and RGB stars are well reproduced by the model. We note
that the apparent overabundance of stars in the brightest bins is due
to the observed LF including AGB stars, while the model LF does not.

As is apparent, this theoretical LF is in excellent agreement with the
empirical one, when allowance is made for the HB and AGB
contributions. The sharp drop near $J=18$ corresponds to the beginning
of the post-MS evolution, and its location is age dependent. However,
no tight limits on the age can be placed here, as this drop would be
displaced towards fainter luminosities by only $\sim 0.1$ mag per 1
Gyr increase in age. The age of the bulge stellar population is
discussed by Ortolani et al. (1995). It will be examined again using a
variety of data now available for this $-6^\circ$ field (Zoccali et
al. 1999).

\subsection{Expected vs Observed Number of Stars}

Having determined the actual slope of the IMF, we are now in a
position to check the theoretical prediction concerning the number of
MS stars in the observed field:
\begin{equation}
N\simeq \int_{0.15}^1 \Psi(M) dM = A \int_{0.15}^1 M^{-1.33} dM
\end{equation}
where $A\simeq 1.2 L_{\rm T}$ (Renzini 1998). From an optical CMD
(Zoccali et al. 1999) referring to a field of $\sim 66$ arcmin$^2$ and
using the distance and reddening adopted in the present paper, we
determine an average surface brightness of the $-6^\circ$ field of
0.55 $\lsun_{\rm V}\,\rm arcsec^{-2}$.  Correcting for the disk
contamination leaves an average surface brightness of the bulge alone
of 0.35 $\lsun_{\rm V}\,\rm arcsec^{-2}$. Hence, the total luminosity
sampled by our 506 arcsec$^2$ field is $\sim 177\,\lsun_{\rm V}$, 
or $L_{\rm T} = 283 \lsun_{\rm Bol}$. Correspondingly, the number of 
stars in the $0.15-1\,\msun$ range is given by:
\begin{equation}
N\simeq 1.2\times 283\int_{0.15}^1 M^{-1.33}dM = 898 \quad {\rm stars}
\end{equation}
which compares well to the 820 stars observed.


\subsection{The $M/L$ Ratio of the Galactic Bulge}
\label{ml}

By integrating the LF shown in Figure~\ref{lf_fro} we determine the
total $J$ band luminosity sampled by a 506 arcsec$^2$ field
to be $L_J=688.5L_{J\odot}$. Note that the stellar population sampled by
our NIC2 field would not be representative of the entire bulge, being
very small, and chosen to be in a region lacking very bright stars. 
However, the bright part of the LF was derived using stars in a 
field 8 times wider, and therefore we can trust the total luminosity
calculated above as representative of the average surface brightness
of the bulge at $b=-6^\circ$.

The total bulge mass in stars included in our 506 arcsec$^2$
field corresponds to the sum of the mass of the
detected stars, plus the mass of M dwarfs and brown dwarfs with
$M<0.15M_\odot$, plus the mass of white dwarfs, neutron stars, and
black hole remnants, the end products of now defunct stars with
$M\gsim1\,\msun$.

We estimate the total mass in our field as follows. First, we simply
sum the masses of the stars actually observed in the field (corrected
for incompleteness and disk contamination) and obtain $317M_\odot$.
By extrapolating the $\alpha=-1.33$ IMF from $M=0.15M_\odot$ all the
way down to zero mass, we obtain $123.5 M_\odot$ of unseen dwarfs,
thus totaling $440.5 M_\odot$ in living stars and brown dwarfs.  To
account for the remnants we need to adopt an initial mass-final mass
relation.  We used the semi-empirical relation proposed by Renzini \&
Ciotti (1993), with white dwarf remnants of mass $M_{\rm
WD}=0.48+0.077\,M_{\rm i}$ for initial masses $M_{\rm i}\le 8\,\msun$,
neutron star remnants of $1.4\,\msun$ for $8\le M_{\rm i}\le
40\,\msun$, and black holes remnants of mass $0.5\,M_{\rm i}$ for
$M_{\rm i}>40\,\msun$.  Since the present data do not give any
constraint on the slope of the IMF for $M\gsim 1\,\msun$, we explore
the effect on the total mass of various plausible assumptions:
\begin{itemize}
\item
IMF \#1:
An IMF with slope $\alpha=-1.33$, like the one we observed, all the
way to $100M_\odot$.  This is perhaps an extreme possibility, since
all the determinations of the IMF in this mass range give steeper
values (see Scalo 1998 for a recent review), and even our own IMF may
steepen for $M>0.5M_\odot$.
\item
IMF \#2:
An IMF with slope $\alpha=-1.33$ up to $M=1M_\odot$ and $\alpha=-2$ for 
$M>1M_\odot$. This is the most conservative assumption, since the 
IMF we observed is best fit with a slope $\alpha=-2$ for $M>0.5M_\odot$. 
\item
IMF \#3:
An IMF with $\alpha=-1.33$ up to $M=1M_\odot$ and $\alpha=-2.35$
(Salpeter's value) for $M>1M_\odot$.
\item
IMF \#4:
Finally, we consider an IMF with $\alpha=-1.33$ up to $M=1M_\odot$,
Salpeter slope for $1<M/M_\odot<2$ and $\alpha=-2.7$ (Scalo 1986) for
$M>2M_\odot$. 
\end{itemize}
For each of these four choices, Table 3 gives the total mass in the
506 arcsec$^2$ field, as well as the contribution
of white dwarfs, neutron stars, and black holes. Of course, the mass
of the unseen dwarfs ($M<0.15\,\msun$) and detected MS dwarfs is the
same for all the IMF options.  Finally, the last column gives the
corresponding $M/L_{\rm J}$ ratio.

Option 1 is clearly top heavy, with most of the bulge baryonic mass in
20-50 $\msun$ black holes. With option 2 one gets rid of most of black
holes, and the mass-to-light ratio drops to near unity. Further
steepening the IMF, such as in options 3 and 4, ceases to have a major
effect on the mass-to-light ratio, while reducing to just a trace
contribution the mass of relativistic remnants. As the microlensing
statistics improves, microlensing experiments may eventually allow us to
select the best among these (or other) options.


\subsection{Gravitational microlensing}
\label{microl}

In this section we consider the implications of our bulge IMF for the
interpretation of the microlensing events that have been observed in
the direction of the Galactic bulge.  Only about 50 of these events
have been published so far (Udalski et al.\ 1994; Alcock et al.\
1997), but by now at least ten times more events should have been
detected.  The initial results have generated two somewhat orthogonal
puzzles.  First, the distribution of event timescales $t_{\rm E}$ is
peaked toward much lower values ($t_{\rm E}\sim 10\,$days) than would
be expected if the bulge IMF were as flat as $\alpha=-0.56$\footnote{
This value is slightly different from $\alpha=-0.54$ quoted in \S\ 7,
because the latter was obtained on the restricted mass interval in
common with our NICMOS data}, as reported by Gould et al.\ (1997) for
the disk MF, but would be well explained by a power-law IMF with
$\alpha\sim-2$ and cut off near the hydrogen-burning limit (Zhao et
al.\ 1995; Han \& Gould 1996). Lower-mass lenses produce events that
on average are shorter ($t_{\rm E}\propto M^{1/2}$), so a steeper IMF
gives rise to a $t_{\rm E}$ distribution skewed toward shorter
timescales.  The slope reported here ($\alpha\sim -1.33$) is
apparently not quite steep enough, though correcting for binaries may
steepen the slope by a few tenths.  Moreover, it has been shown that
many of the shorter events seen toward the bulge are ``amplification
biased'' events of faint sources that are below the threshold of
detection Han (1997).  These are mistaken for events of much brighter
sources in the same seeing disk in which they are detected, and so the
observed timescale for the period of significant apparent
magnification is much shorter than the actual event timescale.  Thus,
the combination of our steeper IMF and the amplification bias may well
allow the bulge microlensing events to be explained by ordinary stars
(perhaps with a smooth extension into the brown dwarf regime).

It is not worth trying here to expand further on the implications of
the bulge IMF for the interpretation of microlensing experiments,
given the very small number of published events compared to the huge
number that will soon become available.  It will then be possible to
make a detailed comparison between the observed timescale distribution
from a large, very clean sample and that predicted on the basis of the
IMF reported here.  As for the $M/L$ ratio, the $f(t_{\rm E})$
distribution will depend not only on the IMF of still living stars,
but also on the number and mass of the dead remnants, white dwarfs,
neutron stars, and black holes. Such a distribution could provide
constraints on the bulge IMF at masses greater than the present
turnoff ($M\sim M_\odot$), even to $\sim 8\,M_\odot$ and beyond (Gould
1999)


\section{Conclusions}
\label{concl}

We have presented the results of stellar photometry on deep images
obtained with NICMOS on board of HST. The data refer to a field in the
Galactic bulge, at a projected distance from the Galactic center of
$\sim 6^\circ$.  From the $J$-band LF of the stars in the field we
derive the IMF of the Galactic bulge with the aid of a theoretical
mass-luminosity relation which provides an excellent fit to the
empirical MLR. The IMF so obtained refers to the mass range from $\sim
1\,\msun$ down to $\sim 0.15\,\msun$, being therefore the deepest IMF
so far obtained for a Galactic bulge. Nevertheless, this low-mass limit
is still nearly a factor of $\sim 2$ above the hydrogen burning limit.

The IMF is well fit by a single slope power law with $\alpha
=-1.33\pm0.07$, therefore much flatter than Salpeter's IMF with
$\alpha=-2.35$.  A two-slope IMF with $\alpha=-2.00\pm 0.23$ above
$0.50\,M_\odot$ and $\alpha=-1.43\pm 0.13$ below gives a better fit,
formally at the $3\,\sigma$ level.  However, in view of the larger
error bars in the upper mass range, and the evolutionary effect away
from the zero-age MS, we prefer to quote the single-slope power law as
our main conclusion. This result is robust within current
uncertainties in the reddening, distance modulus of the Galactic
center, disk and binary stars contamination, and average metallicity
of the bulge stars.

For the mass range in common ($0.35\msun\,\lsim M\lsim 1\msun$), the
derived IMF is in very good agreement with the bulge IMF obtained from
optical observations with WFPC2 by Holtzman et al. (1998).  Our bulge
IMF, however, is appreciably steeper than the low mass IMF for the
solar neighborhood found in two recent determinations, which give
slopes of $\alpha=-0.8$ (Reid \& Gizis 1997) and $\alpha=-0.54$ (Gould
et al. 1997). However, the present bulge IMF is virtually identical to
yet other determinations of the solar neighborhood IMF (Kroupa et
al. 1993, Reid et al. 1999), and an assessment as to whether bulge and
disk IMFs are the same or not will require an understanding the origin
of the large discrepancies among the various determinations of the
disk IMF.

We have also compared the bulge IMF with the present day MF of some
Galactic globular clusters with different metallicities and affected
to various degrees by dynamical processes. In all clusters the MF is
flatter than that of the bulge, but it appears to be closer to the
bulge IMF in those clusters that are less affected by dynamical
processes.  This suggests little or no dependence of the IMF on
metallicity for old systems.

One major issue concerns the amount of bulge mass that is locked in
unseen dwarfs. There is no hint for the IMF slope to change towards
the lower mass limit ($0.15\,\msun$) of the explored range. Assuming
the slope can be extrapolated all the way to mass zero, gives a total
mass of brown dwarfs ($0<M<0.08\,\msun$) in the NIC2 field of
$81\,\msun$, i.e., $<14\%$ of the total stellar mass in the field
(c.f. Table~\ref{tab_ml}).  We note that some support for this
extrapolation comes from the local density of L dwarfs (Reid et
al. 1999).  With all stars making up to $\sim 10\%$ of the total
baryonic mass of the universe, this result suggests that brown dwarfs
may represent not more than $1.4\%$, i.e. a minor fraction, of the baryonic
mass of the universe.

Finally, we have estimated the $M/L_J$ mass to light ratio of the bulge
to be very close to unity ($\sim 0.9\pm0.1$) for reasonable assumptions
of the  IMF slope outside the directly explored range.


\acknowledgments

We are expecially grateful to Claudia Maraston for help in constructing 
Table~\ref{tab_ml}. We also thank Peter Stetson for providing us the
software to calculate the stellar PSF. We finally thank the anonymous 
referee for the many constructive comments that have significantly 
improved the paper.
Support for this work was provided by NASA through grant number GO-7891
from the Space Telescope Science Institute , which is operated by the
Association of Universities for Research in Astronomy, Incorporated, under
NASA contract NAS5-26555.



\newpage

\begin{deluxetable}{rrrrc}	
\tablewidth{20pc}
\tablenum{1}
\tablecaption{The Luminosity Function \label{tab_lf}}
\tablehead{
\colhead{$J$} &
\colhead{$N$} &
\colhead{$N_c$} &
\colhead{$\sigma$} &
\colhead{Disk frac. } 
} 
\startdata
   16.5 &  5 &   5 &   2  & 0.20 \\  
   17.0 &  7 &   7 &   3  & 0.18 \\  
   17.5 &  6 &   6 &   2  & 0.18 \\  
   18.0 & 20 &  18 &   4  & 0.14 \\  
   18.5 & 18 &  19 &   4  & 0.13 \\  
   19.0 & 25 &  26 &   5  & 0.13 \\  
   19.5 & 32 &  32 &   6  & 0.13 \\  
   20.0 & 38 &  40 &   6  & 0.13 \\  
   20.5 & 45 &  50 &  10  & 0.12 \\  
   21.0 & 54 &  60 &  10  & 0.12 \\ 
   21.5 & 69 &  82 &  15  & 0.11 \\ 
   22.0 & 66 &  81 &  14  & 0.11 \\  
   22.5 & 86 & 112 &  19  & 0.10 \\  
   23.0 & 81 & 119 &  22  & 0.10 \\ 
   23.5 & 73 & 119 &  21  & 0.10 \\ 
   24.0 & 65 & 139 &  31  & 0.09 \\  
   24.5 & 57 & 162 &  27  & 0.08 \\  
\tableline
\enddata
\end{deluxetable}

\begin{deluxetable}{rrrc}  
\tablewidth{20pc}
\tablenum{2}
\tablecaption{Bright extension of the LF \label{tab_disk}}
\tablehead{
\colhead{$J$} &
\colhead{$N$} &
\colhead{$\sigma$} &
\colhead{Disk frac.} 
} 
\startdata
    8.750  &  0.010  & 0.002	&  0.00	\\
    9.250  &  0.009  & 0.003	&  0.00	\\
    9.750  &  0.009  & 0.003	&  0.00	\\
   10.250  &  0.026  & 0.004	&  0.00	\\
   10.750  &  0.033  & 0.004	&  0.00	\\
   11.250  &  0.052  & 0.005	&  0.00	\\
   11.750  &  0.032  & 0.005	&  0.00	\\
   12.250  &  0.112  & 0.116	&  0.00	\\	
   12.750  &  0.112  & 0.113 	&  0.05	\\
   13.250  &  0.661  & 0.268	&  0.10	\\
   13.750  &  0.447  & 0.223	&  0.08	\\
   14.250  &  0.661  & 0.263	&  0.13	\\	
   14.750  &  0.661  & 0.246	&  0.24	\\
   15.250  &  1.000  & 0.290	&  0.30	\\
   15.750  &  1.230  & 0.310	&  0.35	\\
   16.250  &  1.660  & 0.343	&  0.41	\\
   16.750  &  2.754  & 0.464	&  0.35	\\
\tableline
\enddata
\end{deluxetable}

\begin{deluxetable}{cccccccc}	
\tablewidth{40pc}
\tablenum{3}
\tablecaption{Bulge $M/L_J$ ratios \label{tab_ml}}
\tablehead{
\colhead{$\alpha$} &
\colhead{$\alpha$} &
\colhead{$\alpha$} &
\colhead{$M_{TOT}$} &
\colhead{$M_{WD}$} &
\colhead{$M_{NS}$} &
\colhead{$M_{BH}$} &
\colhead{$M/L_J$} \nl
\colhead{$M<1M_\odot$} &
\colhead{$1<M/M_\odot<2$} &
\colhead{$M>2M_\odot$} & & & & &
} 
\startdata
$-1.33$ & $-1.33$ & $-1.33$ & 3207 & 326 & 247 & 2213 & 4.7 \\
$-1.33$ & $-2.00$ & $-2.00$ &  744 & 150 &  38 &  135 & 1.1 \\
$-1.33$ & $-2.35$ & $-2.35$ &  601 & 133 &  15 &   32 & 0.9 \\
$-1.33$ & $-2.35$ & $-2.70$ &  562 & 124 &   8 &   10 & 0.8 \\
\tableline
\enddata
\end{deluxetable}

\end{document}